\documentclass[fleqn,usenatbib]{mnras}

\usepackage{newtxtext,newtxmath}

\usepackage[T1]{fontenc}
\usepackage{ae,aecompl}

\usepackage{graphicx,subcaption}	
\usepackage{float}
\usepackage{caption}
\usepackage{amsmath}	
\usepackage{amssymb}	
\usepackage{amsfonts}

\title[High-cadence observations of Swift J1818.0$-$1607]{High-cadence observations and variable spin behaviour of magnetar Swift J1818.0$-$1607
after its outburst}

\author[Champion et al.]{David Champion$^{1}$, 
Ismael Cognard$^{2,3}$,
Marilyn Cruces$^1$,
Gregory Desvignes$^{1,4}$, 
\and
Fabian Jankowski$^{5}$,
Ramesh Karuppusamy$^{1}$,
Michael~J.~Keith$^{5}$\thanks{Contact author: mkeith@pulsarastronomy.net}, 
Chryssa Kouveliotou$^{6,7}$,
\and
Michael Kramer$^{1,5}$,
Kuo Liu$^1$\thanks{Contact author: kliu@mpifr-bonn.mpg.de},
Andrew G. Lyne$^{5}$, 
Mitchell B. Mickaliger$^{5}$, 
\and
Brendan O'Connor$^{6,7}$,
Aditya Parthasarathy$^1$,
Nataliya Porayko$^{1}$,
 Kaustubh Rajwade$^{5}$,
 \and
 Ben W. Stappers$^{5}$,
 Pablo Torne$^{8,9}$,
 Alexander J. van der Horst$^{6,7}$
 Patrick Weltevrede$^{5}$
\\
\\
$^{1}$Max Planck Institute for Radio Astronomy, Auf dem H\"ugel 69, Bonn D-53121, Germany
\\
$^2$Station de Radioastronomie de Nançay, Observatoire de Paris, CNRS/INSU, Université d'Orléans, 18330, Nançay, France \\
$^3$Laboratoire de Physique et Chimie de l'Environnement, CNRS, 3A Avenue de la Recherche Scientifique, 45071, Orléans Cedex 2, France 
\\
$^4$LESIA, Observatoire de Paris, Université PSL, CNRS, Sorbonne Université, Université de Paris, 5 Place Jules Janssen, 92195, Meudon, France
\\
$^{5}$ Jodrell Bank Centre for Astrophysics, University of Manchester, M13 9PL Manchester, UK
\\
$^6$Department of Physics, the George Washington University, 725 21st Street NW, Washington, DC 20052, USA
\\
$^7$ Astronomy, Physics, and Statistics Institute of Sciences (APSIS), The George Washington University, Washington, DC 20052, USA
\\
$^8$ Instituto de Radioastronom\'ia Milim\'etrica (IRAM), Avda. Divina Pastora 7, Local 20, 18012 Granada, Spain
\\
$^9$ East Asian Observatory, 660 N. A’ohoku Place, University Park, Hilo, Hawaii 96720, USA
}

\date{Accepted XXX. Received YYY; in  original form ZZZ}

\pubyear{2020}

\begin{document}
\label{firstpage}
\pagerange{\pageref{firstpage}--\pageref{lastpage}}
\maketitle
\graphicspath{{plots/}}

\begin{abstract}
We report on multi-frequency radio observations of the new magnetar Swift J1818.0$-$1607, following it for more than one month with high cadence. The observations commenced less than 35 hours after its registered first outburst. We obtained timing, polarisation and spectral information. Swift J1818.0$-$1607 has an unusually steep spectrum for a radio emitting magnetar and also has a relatively narrow and simple pulse profile. The position angle swing of the polarisation is flat over the pulse profile, possibly suggesting that our line-of-sight grazes the edge of the emission beam. This may also explain the steep spectrum. The spin evolution shows large variation in the spin-down rate, associated with four distinct timing events over the course of our observations. Those events may be related to the appearance and disappearance of a second pulse component. The first timing event coincides with our actual observations, while we did not detect significant changes in the emission properties which could reveal further magnetospheric changes. Characteristic ages inferred from the timing measurements over the course of months vary by nearly an order of magnitude. A longer-term spin-down measurement over approximately 100 days suggests an characteristic age of about 500 years, larger than previously reported. Though Swift J1818.0$-$1607 could still be one of the youngest neutron stars (and magnetars) detected so far, we caution using the characteristic age as a true-age indicator given the caveats behind its calculation.
\end{abstract}



\section{Introduction}
Among the population of young Galactic neutron stars, magnetars form their own class \citep[for an overview, see][]{kb17}. Most of them are persistent X-ray sources, however, several have been also observed at optical, IR, and radio wavelengths. 
Their most characteristic attributes are the emission of repeated, very short ($\sim100$\,s of ms) hard X-ray bursts during randomly occurring outbursts, and an X-ray luminosity higher than the spin-down luminosity.
Of the $\sim30$ magnetars known to date, only two have been located outside the Milky Way: one each in the Large and Small Magellanic Clouds. 
Their periods, $P$, range between $\sim$1 and 12 seconds \citep{ok14}\footnote{\url{http://www.physics.mcgill.ca/~pulsar/magnetar/main.html}} and their period derivatives, $\dot{P}$, imply magnetic field strengths, computed via $B=3.2\times 10^ {19} \sqrt{P \dot{P}}$, typically of the order of $10^{14} - 10^{15}$ Gauss \citep{cds+98}. 
Magnetar emission is thus believed to be powered by their strong magnetic fields causing neutron star quakes and magnetospheric phenomena, giving rise to the designation of these objects \citep{dt92}. 
%
Using the expression for the ``characteristic age'' derived for radio pulsars, $\tau = P/2\dot{P}$, one finds that magnetars are expected to be young, with typical ages of a few hundred to a few thousand years. Consequently, magnetars occupy the upper right corner of the ``$P-\dot{P}$ diagram'', relative to radio pulsars. We refer to recent reviews for a detailed discussion of magnetar properties (e.g.~\citealp{kb17}) or their relationship to pulsars and other neutron star populations (e.g.~\citealp{kk15}).

Only a handful of magnetars have been detected at radio frequencies. The first radio detection of XTE J1810$-$197 \citep{crh+06} revealed a number of characteristics of radio-loud magnetars that are archetypal.
Firstly, magnetar radio emission is transient. It sometimes appears to be connected to high-energy outbursts \citep{hgb+05}, though the behavior is not consistent \citep{lld+19}. In the unusual case of PSR J1622$-$4950, the magnetar was first discovered via its radio emission while its X-ray emission was in quiescence \citep{lbb+10}. 
The radio-loud magnetar in the Galactic Centre, PSR J1745$-$2900 \citep{efk+13}, is probably the source with the longest duration of detectable radio emission known since its initial outburst in 2013, albeit with significant variations in its radio brightness (e.g., \citealt{dep+18}). 

All radio-detected magnetars share a very large degree of linear polarisation in their emission, both in the single pulses and in the average profile (e.g. \citealp{ksj+07,crj+07,lbb+10,efk+13,dai2019}), even up to frequencies as high as 150 GHz \citep{tor17}. 
The most detailed polarisation study thus far was conducted by \cite{ksj+07}, using simultaneous multi-frequency observations of single pulses and the average pulse profile at several epochs of XTE J1810$-$197. They found that the emission is nearly 80-95 per cent linearly polarised, often with a low but significant degree of circular polarization at all frequencies. This can be even greater in selected single pulses. The position angle is typically difficult to interpret, especially since it is shown to change with time. This variability of polarised emission was confirmed when XTE J1810$-$197 recently switched on as a radio source again ten years after the cease of its radio emission in 2008 \citep{lld+19,dai2019}. 

The individual pulses of magnetars are also commonly very narrow (`spiky'), spreading over an often wide pulse window \citep{ccr+07,ksj+07,tor15}, and longitude-resolved modulation indices reveal a high degree of intensity fluctuations on day-to-day timescales and dramatic changes across pulse phase \citep{ssw+09}. This variability in the individual pulses is also reflected in the varying shapes of average pulse profiles \citep{ccr+07,lld+19}. The `spikiness' combined with the large variability, and the high degree of polarisation, resembles some of the properties of repeating Fast Radio Bursts (FRBs) \citep{ssh+16}, and so it is not surprising that magnetars are among the models to explain FRBs \citep[e.g.][]{lyu14,maan2019}.

Unlike normal pulsars, the radio emission of magnetars typically shows a flat spectrum \citep{crh+06,ssw+09,tor15,dai2019}, with variation in the single pulse spectra \citep{ssw+09,tor15} that is larger than that for normal pulsars (e.g.~\citealp{kkg+03}). The flatness of the radio spectrum has led to the detection of radio-loud magnetars up to a frequency of $\sim$300\,GHz \citep{tor17, tor2020}, which is the highest radio frequency of any neutron star detection so far.

It is intriguing to contrast the radio emission of magnetars with that of `normal' radio pulsars. Overall, as the above brief description shows, one finds that in several respects magnetar emission shows similarities to the emission properties of normal radio pulsars while simultaneously showing striking differences \citep{ksj+07}. Understanding the extent, and potentially the origin, of these differences and similarities promises to help solve the radio emission mystery of radio emitting neutron stars as a whole. Extending these studies with additional, new radio-loud magnetars is therefore extremely useful. 

Determining the relationship between rotation and magnetic powered neutron stars is also important in understanding the formation of magnetars and population of neutron stars as a whole. As pointed out by \cite{kk08}, the neutron star birthrate and population estimates are not consistent with the Galactic supernova rate. This problem would be alleviated if one considers a possible evolutionary scenario between some of the known neutron star classes. The possibility of such a scenario was demonstrated by \cite{elk+11}, who pointed out that PSR J1734$-$3333, a radio pulsar rotating with a period $P = 1.17$ s and slowing down with a period derivative $\dot{P}=2.28\times 10^{-12}$, is located in the `$P$-$\dot{P}$-diagram' midway between those of normal rotation-powered pulsars and magnetars. In case of an unchanged braking index, this pulsar may soon have the rotational properties of a magnetar. The existence of such an evolutionary channel is supported by a few other cases such as PSR~J1119$-$6127 and PSR~J1846$-$0258 which also appear to be someway between rotation powered pulsars and magnetars \citep{ggg+08,djw+18}.

One of the latest additions to this small sample of magnetars that can be studied both in terms of radio emission and spin-evolution is Swift J1818.0$-$1607 \citep{egk+2020}. Following the Swift/BAT detection of a short burst from this source, \citet{esy+2020} reported the detection of coherent X-ray pulsations with a period of 1.36 s, and suggested that Swift J1818.0$-$1607 is a new magnetar. With a spin period of 1.36\,s it would be at least the second shortest among the known magnetars\footnote{The magnetar-like pulsar PSR~J1846$-$0258, though having a spin period of approximately 0.33\,s, might be from a distinct population of high-B rotation-powered pulsars \citep{ggg+08,lnk+11,khd18}}. About 35 hours after the X-ray burst, observations with the 100-m Effelsberg telescope discovered radio pulsations from Swift J1818.0$-$1607 \citep{kdk+20}, which were soon confirmed by independent observations with the Lovell Telescope \citep{rsl+20} and, later, by observations with other telescopes that we refer to later in the paper.

The initial Effelsberg observations already established that the radio pulses exhibited a very high degree of linear polarisation, consistent with the magnetar emission properties described above and in agreement with \citet{lsjb2020}, giving support to the suggested magnetar-nature. The confirmation that Swift J1818.0$-$1607 is indeed a magnetar was provided by timing observations by the Effelsberg and Lovell telescopes: \cite{cdj+20} presented the first measurement of a period derivative of $\dot{P}= 8.16(2) \times 10^{-11}$ and, hence, a derived characteristic age of 265(1) yrs assuming a constant (since birth) braking index of 3 and a birth period much less than the current period. This measured spin-down also implies a characteristic surface dipole magnetic field of $B=3.4 \times 10^{14}$ G and a spin-down luminosity of $\dot{E}=1.3 \times 10^{36}$ erg s$^{-1}$ (assuming a moment of inertia of $10^{38}$ kg m$^2$). These values confirmed the magnetar nature of the detected source. As pointed out by \cite{cdj+20}, the characteristic age is very similar to that of SGR 1806$-$20 which currently has the smallest characteristic age on record (240 years; \citealp{ok14}). The measurement of spin-down rate was later confirmed by X-ray timing \citep{nicerglitch,esposito20} which also revealed an early timing anomaly \citep{nicerglitch}.

In the following, we present continued, extensive radio observations of Swift J1818.0$-$1607, performed with high cadence with the Effelsberg 100-m, the Lovell Telescope at Jodrell Bank Observatory (JBO) and the Nan\c{c}ay Radio Telescope (NRT). We mostly concentrate on the first month after the magnetar activation, but will also present results from continuing observations. This comprehensive study reveals, among other findings, a complex timing behaviour and significant pulse-shape changes. As we show, a short-term assessment of the spin-down evolution can bias the derived magnetar properties significantly. We also present a summary of updated and new results for the emission properties of the source.

\section{Observations}
\label{sec:obs}

 The results reported here come from observations with the Effelsberg 100-m, the Lovell 76-m and the Nan\c cay radio telescopes and are summarised in Table \ref{dataset_table}. While observations continued to the time of writing, this work covers the first month since the magnetar's outburst, i.e. from 2020 March 14 to 2020 April 15. 
 
 At Effelsberg, Swift J1818.0$-$1607 was observed 28 times starting from the first observations carried out at 1.37 GHz on 2020 March 14, with the subsequent observations at 1.37, 2.55, 4.85 or 6 GHz (see Table \ref{dataset_table} for details).
 Signal processing for the measurements at 1.37, 2.55 and 4.85 GHz involved real-time coherent dedispersion followed by averaging in to sub-integrations of 10$\,$s duration starting from the second session using the PSRIX pulsar backend \citep{lkg+2016}. Where possible, additional baseband data were recorded with 320, 400 or 500 MHz bandwidth for the three different receivers used. These data were processed offline to generate both single pulses and 10$\,$s sub-integrations after coherent dedispersion and folding. We used dspsr \citep{vsb2011} to coherently dedisperse and fold the data in both real-time and offline processing pipelines. Observations at 6 GHz with the wide band receiver covering 4-8 GHz are recorded in search mode with a time and frequency resolution of 131 $\mu$s and $\sim 1$ MHz, respectively \citep[see e.g.][]{dep+18}. In each session, we recorded a 1\,Hz switched noise diode for 90--120\,s to allow for polarisation calibration. The calibration was conducted for frequency-dependent gain and phase variations between the signals received by the pulsar instrument from the two orthogonal probes of the receiver. Subsequently, Faraday rotation was measured and corrected to reveal the polarisation profile. In addition, the data were manually checked to remove apparent radio interference (RFI). The software package \textsc{psrchive} \citep{vdo12} was used for most of the data post-processing. 

At JBO, the source was observed with the Lovell Telescope at a centre frequency of 1.53 GHz,  with 512 MHz of bandwidth divided into 1532 channels, over four days, starting on 2020 March 14. Complex voltage data from the telescope were converted into Stokes parameters using a ROACH-1 Field Programmable Gate Array board. Then, they were processed by the DFB backend~\citep{manchester2013}, folding the data modulo the topocentric period and dedispersing them at the dispersion measure (DM) reported in \citet{kdk+20}. The resulting time-frequency data were folded using the best ephemeris into multiple 8-second sub-integrations and saved to disk.
The data were subsequently visually inspected and manually cleaned to remove strong radio frequency interference (RFI) using the \textsc{psrchive} package, leaving approximately 250 MHz of usable bandwidth.
To calibrate the polarisation of the JBO data, we used RFI-cleaned datasets of a bright pulsar with well known polarization properties that was observed close in time to each of our magnetar observations as a template for the polarization calibration. We generated the receiver solutions for the Lovell Telescope signal chain using the Measurement Equation Template Matching (METM) method~\citep{vanstraten13} by solving for the receiver solution of the bright pulsar data, namely PSR B0919+06 using a fully calibrated data set of the same pulsar obtained from the EPN pulsar database~\citep{johnston2020}. Under the assumption that there is no significant change in the signal chains we then use the receiver solutions to calibrate the magnetar observations. We fit for the differential gain and phase as a function of time to obtain the best fit polynomial for the receiver solution using the publicly available software \textsc{PSRSALSA}~\citep{weltevrede16}. Finally, we corrected the PA for Faraday rotation using the value of RM obtained from the Effelsberg data. These have led high consistency with the Effelsberg polarisation properties (see Section~\ref{sssec:pol} for more details).

At the NRT, the source was observed on seven epochs from 2020 March 28 to 2020 April 15 at a central frequency of 1.48 GHz. The Nan\c{c}ay Ultimate Pulsar Processing Instrument (NUPPI) backend was used to record the data in PSRFITS\footnote{\url{https://www.atnf.csiro.au/research/pulsar/psrfits_definition/PsrfitsDocumentation.html}} search mode format, with 512\,MHz bandwidth and 4-MHz frequency resolution. The data were then folded offline using the pulsar ephemeris determined from observations before 2020 March 28, into 10-s sub-integrations. Each sub-integration was visually inspected and manually cleaned to remove RFI using the \textsc{psrchive} software package, which leaves an overall effective bandwidth of approximately 420\,MHz.

\begin{table*}
    \centering
    \caption{Details of the observations carried out during the first month after the magnetar activation.}

    \begin{tabular}[c]{cccccc}
\hline\hline
\textbf{UTC start} & \textbf{Duration (min)} & \textbf{Telescope} & \textbf{Centre Frequency (GHz)} & \textbf{Bandwidth (MHz)} & \textbf{Details } \\
\hline
2020-03-14 06:37 &60 &EFF  &1.370 & 240 & Single component \\
2020-03-14 10:24 &36 &JBO  &1.532 & 336 & Single component \\
2020-03-15 05:20 &60 &JBO  &1.532 & 336 & Single component \\
2020-03-15 07:25 &31 &EFF  &2.550 & 130 & Single component \\
2020-03-15 07:25 &69 &JBO  &1.532 & 336 & Single component \\
2020-03-15 09:43 &60 &JBO  &1.532 & 336 & Single component \\
2020-03-16 03:22 &30 &JBO  &1.532 & 336 & Single component \\
2020-03-16 10:03 &30 &JBO  &1.532 & 336 & Single component \\
2020-03-17 03:57 &85 &JBO  &1.532 & 336 & Single component \\
2020-03-17 04:17 &70 &EFF  &4.850 & 500 & Not detected \\
2020-03-19 08:37 &32 &EFF  &1.370 & 240 & Double component \\
2020-03-20 06:40 &40 &EFF  &2.550 & 130 & Double component \\
2020-03-20 07:58 &59 &EFF  &1.370 & 240 & Double component \\
2020-03-21 03:25 &322 &EFF  &1.370 & 240 & Double component \\
2020-03-22 03:42 &19 &EFF  &1.370 & 240 & Single component \\
2020-03-22 08:42 &19 &EFF  &1.370 & 240 & Single component \\
2020-03-27 04:09 &20 &EFF  &1.370 & 240 & 7 pulses detected \\
2020-03-27 05:12 &80 &EFF  &2.550 & 130 & 9 pulses detected \\
2020-03-28 05:12 &33 &NRT  &1.484 & 512 & Single component \\
2020-03-29 06:07 &8 &EFF  &6.000 &4000 &Weak detection \\
2020-03-29 06:18 &15 &EFF  &2.550 & 130 & Single component \\
2020-03-29 07:30 &45 &EFF  &1.370 & 240 & Single component \\
2020-03-30 05:21 &38 &NRT  &1.484 & 512 & Single component \\
2020-03-31 06:36 &28 &EFF  &2.550 & 130 & Single component \\
2020-03-31 07:36 &78 &EFF  &1.370 & 240 & Single component \\
2020-04-01 03:49 &45 &EFF  &1.370 & 240 & Single component \\
2020-04-04 02:31 &60 &EFF  &1.370 & 240 & Single component \\
2020-04-04 04:05 &240 &EFF  &2.550 & 130 & Single component \\
2020-04-05 05:23 &120 &EFF  &1.370 & 240 & Single component \\
2020-04-06 04:38 &33 &NRT  &1.484 & 512 & Single component \\
2020-04-06 05:22 &13 &EFF  &6.000 &4000 & Weak detection \\
2020-04-06 05:37 &20 &EFF  &2.550 & 130 & Single component \\
2020-04-08 06:38 &14 &EFF  &1.360 & 140 & Single component \\
2020-04-08 07:21 &32 &EFF  &2.550 & 130 & Single component \\
2020-04-09 04:46 &33 &NRT  &1.484 & 512 & Single component \\
2020-04-10 04:24 &33 &NRT  &1.484 & 512 & Single component\\
2020-04-10 06:05 &27 &EFF  &2.550 & 130 & Single component \\
2020-04-11 06:03 &28 &EFF  &1.370 & 240 & Single component \\
2020-04-13 04:09 &35 &NRT  &1.484 & 512 & Single component\\
2020-04-14 04:48 &69 &EFF  &1.370 & 240 & Single component \\
2020-04-14 07:09 &53 &EFF  &2.550 & 130 & Single component \\
2020-04-15 04:32 &33 &NRT  &1.484 & 512 & Single component\\
\hline
\end{tabular}%
\label{dataset_table}
\end{table*}

\section{Data analysis \& Results}

Preliminary results of the first week of the observations reported here were already presented in \cite{kdk+20}, \cite{rsl+20} and \cite{cdj+20}. Here we provide further data to conduct a comprehensive study of the spin and emission properties of Swift J1818.0$-$1607 during the first month after the magnetar activation.

\subsection{Dispersion measure, rotation measure and pulse scattering}

In the report of the discovery of radio emission, \cite{kdk+20} presented a DM of $706(4)$ cm$^{-3}$pc and a rotation measure (RM) of $+1442\pm3$ rad m$^{-2}$, based on Effelsberg observations at 1.37 GHz. 
The same day, \cite{rsl+20} measured DM$=703\pm7$ cm$^{-3}$pc from JBO observations at 1.53~GHz.
A few days later, \cite{cdj+20} presented a refined DM value of $701(1)$ cm$^{-3}$pc, based on the early observations from both telescopes.

\cite{lowerMeerKAT} determined a DM of 699.5(3) cm$^{-3}$pc, using the MeerKAT telescope at a central frequency of 1.28 GHz, over a 856 MHz bandwidth. This analysis used scatter-broadened templates to account for scattering at the lower edge of the band, and reported a characteristic scattering time of $\tau_{\rm SC, 1 GHz} = 44(3)$ ms at 1 GHz. Later observations with the Parkes telescope derived a similar value of $\tau_{\rm SC, 1 GHz} = 42^{+9}_{-3}$ ms, with a scattering index of $\alpha_{\rm SC}=-3.4^{+0.3}_{-0.2}$ \citep{lsjb2020}.
They also infer a mean DM of $706.0(2)$ cm$^{-3}$pc and $\textrm{RM} = 1442.0(2)$ rad m$^{-2}$ from observations across the 0.8 to 4.0 GHz band.

As indicated by the analysis undertaken by \cite{lowerMeerKAT} the measured DM value may be affected by the presence of interstellar scattering effects. Hence, we have used a code to model the DM and scattering at the same time, while representing the intrinsic pulse profile with Shapelets, similarly to \citet{lkd+17}.
We divided the available 512 MHz-bandwidth NRT data into four subbands and by applying the Shapelet model we obtained a DM value of $701.8(3)$ pc cm$^{-3}$ and a scattering timescale of $\tau_{\rm SC, 1 GHz} = 44(3)$ ms  \citep[assuming the scattering index $\alpha_{\rm SC}$ as determined by][]{lsjb2020}.
Although our scattering timescale agrees with the result reported by \cite{lsjb2020}, we find that our
new DM value is significantly smaller, closer to the value reported by \cite{lowerMeerKAT} using MeerKAT, and consistent with the value by \cite{cdj+20}.
We believe our estimate of the DM is robust and attribute the variation in the reported DM values to different (or absence of) methods of accounting for the bias in DM estimation from scattering, rather than intrinsic variation in the DM.
These DM values correspond to a distance in the range from 4.8 to 8.1 kpc, as determined using the YMW16 \citep{ymw16} and NE2001 \citep{NE2001} electron density models, respectively, and such a large inferred distance is consistent with the observed interstellar scattering.

\subsection{Timing and spin evolution}
\label{sec:timing}

Each observation described in Section~\ref{sec:obs} was averaged in frequency and time to form sub-integrations of approximately 200s in length. As reported in Section \ref{sec:profile}, the radio emission from Swift J1818.0$-$1607 exhibits significant changes in pulse shape over time and so time-of-arrival measurements were made using an adaptive template matching scheme with a variable shape template. For the 1.3-1.5 GHz observations a two-component template was used, with the relative height of the leading component being a free parameter. The width and relative phase of each component was determined using a global fit to a sample of 40 profiles and then held fixed for the processing of the full data set. The 2.5-GHz data were fit using the same template, however an additional component was added to model the bright additional trailing component seen in one observation (2020 March 20) with relative phase fixed using the average profile of that observation. The same template was applied to each instrument, with arbitrary phase jumps included in the timing model to account for any profile evolution between the central frequencies of the instruments. Template fitting was performed using a bespoke python code based on \textsc{psrchive} and fitting was carried out using a Markov-chain Monte-Carlo (MCMC) approach with \textsc{emcee} \citep{2013ForemanMackey}.

\begin{figure*}
    \centering
    \includegraphics[width=0.8\textwidth]{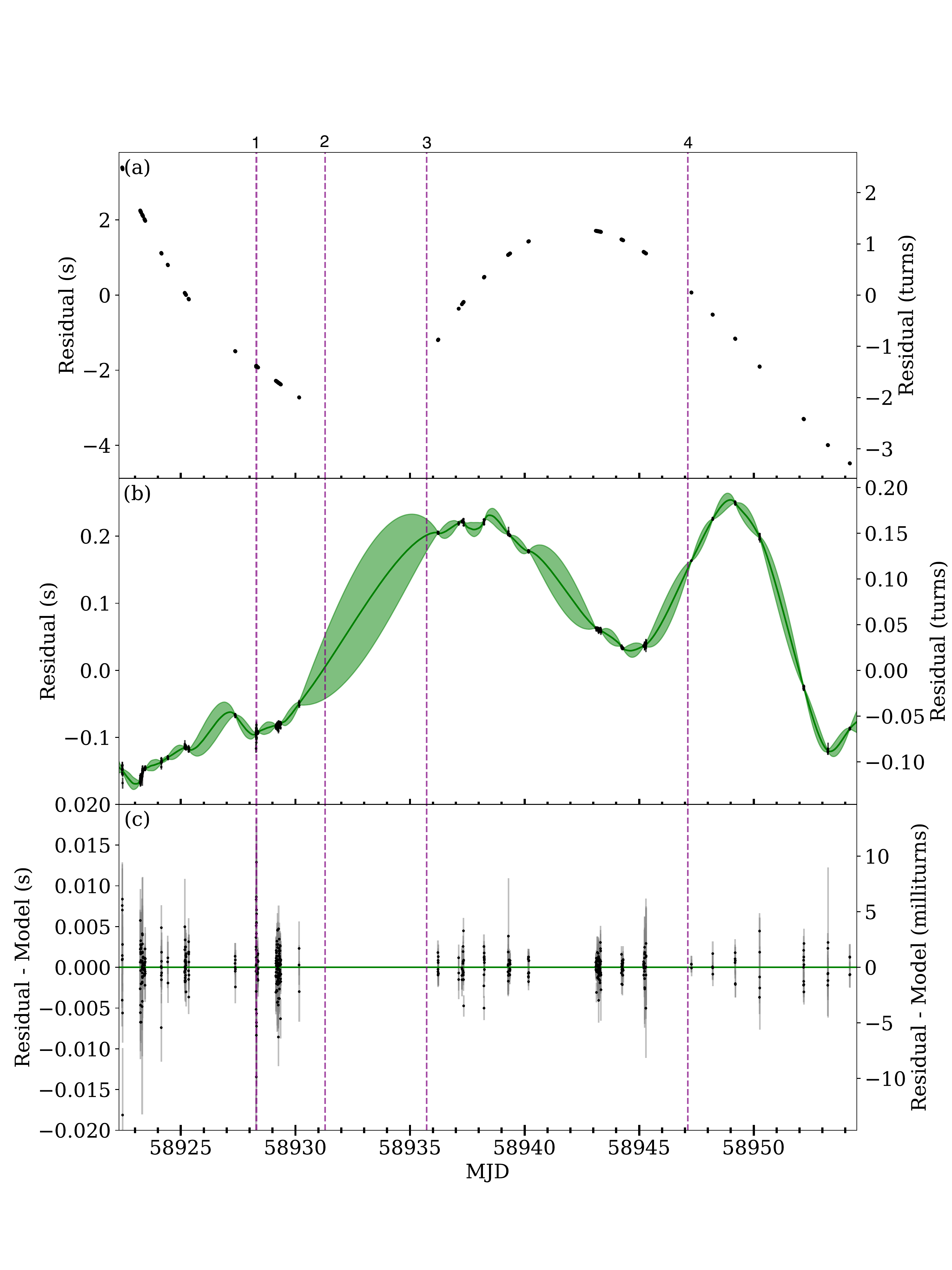}
    \caption{Timing Residuals for Swift J1818.0$-$1607. Upper panel shows the timing residual after fitting only for frequency and frequency derivative. Middle panel shows the residual after subtraction of the step changes in frequency and frequency derivative. The green curve indicates the maximum-likelihood Gaussian process model. Lower panel shows the residual after subtracting the Gaussian process timing model. In all panels the dashed lines show the maximum-likelihood transitions between timing states.}
    \label{fig:timing_residual}
\end{figure*}

The spin frequency of the magnetar evolves significantly over time, changing from $\nu=0.73341\,\mathrm{Hz}$ to $0.73334\,\mathrm{Hz}$ over our observing span of 32 days, resulting in a mean frequency derivative  $\dot{\nu} =  -2.7\times10^{-11}\,\mathrm{Hz^2}$, equivalent to a period derivative $\dot{P} = 5.1 \times 10^{-11}$. However, it is immediately apparent that the spin-down rate also varies significantly over the observing span, ranging between $-1$ and $-6 \times 10^{-11}\,\mathrm{Hz^2}$ when averaged over several day spans, and clear signs are seen of discrete timing {\it events} in the data potentially associated with step changes in spin parameters.
A thorough timing analysis was performed using a pipeline based on \textsc{tempo2} \citep{lah+13} and \textsc{enterprise} \citep{evt+19} software packages to perform a Bayesian timing analysis, with sampling performed using \textsc{emcee}. As with other magnetars, Swift J1818.0$-$1607 exhibits spin noise on a wide range of timescales, and so we include the typical power-law Gaussian process to model the red noise, plus EFAC and EQUAD white noise parameters for each instrument (see \citealp{lah+13} for a description of the noise model). As is common practice, we fit for the index, $\alpha$, and log-amplitude, $\log_{10}(A_\mathrm{red})$, of the power-law Gaussian process such that the power-spectral density of the red noise is given by
\[
P(f) = \frac{A_\mathrm{red}^2}{12\pi^2} \left(\frac{f}{\mathrm{yr}^{-1}}\right)^{-\alpha} \mathrm{yr}^3.
\]
In order to explore the evidence for discrete timing events in the data, we developed a large number of competing models with between 0 and 4 discrete state switches, where the state switches may include step changes in $\nu$, $\dot{\nu}$ and/or $\ddot{\nu}$. The epochs of the transitions between states had uniform priors within large, non-overlapping windows around the suspected events determined by eye. We also explored models including a fixed value of $\ddot{\nu}$ over the observing span, but without step changes in $\dot{\nu}$. Model selection was performed using a trans-dimensional MCMC approach with \textsc{enterprise}, and verified by direct computation of evidence using \textsc{temponest}. The preferred models have 4 discrete changes in $\dot{\nu}$ with or without a step change in $\nu$ at the first transition, and all other models were rejected with an odds ratio of less than 0.01. Table \ref{timing_table} shows the results for the model both with and without a step change in $\nu$ at the first transition, and both the pre-fit and post-fit residuals for the maximum-likelihood model are shown in Figure \ref{fig:timing_residual}. The parameters for the complete maximum-likelihood timing model are given in Table \ref{par_table}. Regardless of the choice of model, the period evolution is well constrained, with only subtle differences in how much of the frequency evolution is attributed to ``noise'' versus ``deterministic'' changes.
Although `by eye' there seems to be a suggestion of cubic behaviour in the residuals, this is likely a result of our subtraction of a quadratic model from the data and the algorithms do not find sufficient evidence to support including the $\ddot{\nu}$ term in the model.

The first timing event at MJD 58928 (2020 March 20) stands out from the others as modelling it also requires a significant change in spin-frequency, $\Delta\nu=1.2(5)$ $\mu$Hz.
This value is somewhat smaller but largely consistent with the value obtained from NICER observations around the same day \citep{nicerglitch}, although the change in spin-down rate determined from our dense radio observations is a few times larger than they reported.

Even when removing the deterministic spin-down model, there is a strong red-noise process remaining in the residuals. From the best-fitting model with 4 discrete state switches we estimate $\alpha = 3.5(3)$ and $\log_{10}(A_\mathrm{red}) = -5.7(3)$ which implies spin variations 3 orders of magnitude larger than the long-term red noise processes seen in normal pulsars \citep{psj+19}, or alternatively that Swift J1818.0$-$1607 experiences a similar amount of timing noise over our 2-month dataset that a typical pulsar might experience in 10 years. For comparison with other studies, we also report these parameters for a model without any $\nu$ and $\dot{\nu}$ changes as $\alpha = 5.0(2)$ and $\log_{10}(A_\mathrm{red}) = -4.0(2)$, although this model is disfavoured in our analysis.

\begin{figure*}
    \centering
    \includegraphics[width=0.8\textwidth]{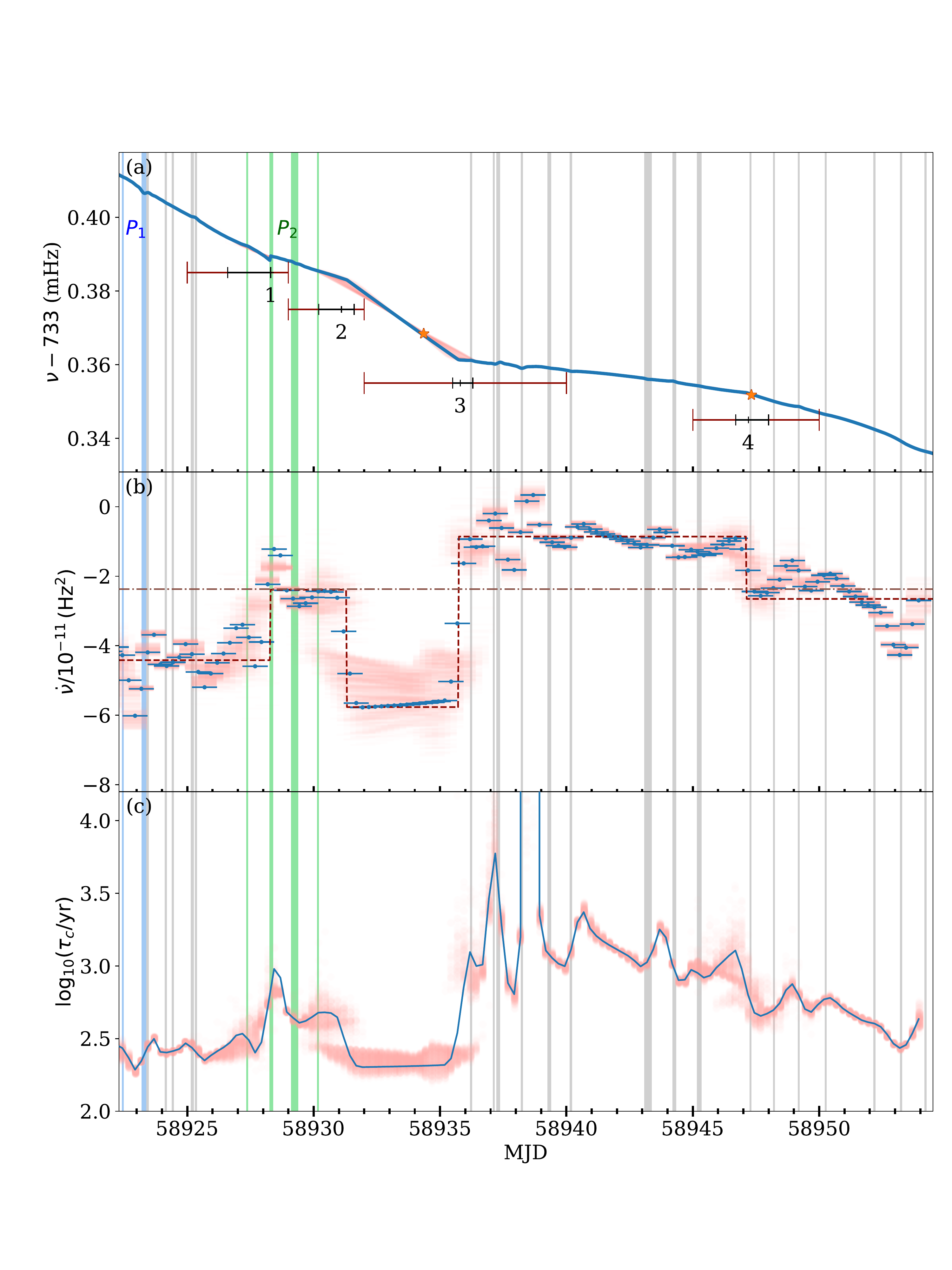}
    \caption{Spin evolution of Swift J1818.0$-$1607. Upper panel shows the modeled frequency evolution over time. Transitions in timing are indicated with numbers, with the prior range shown in red and the posterior range shown in black. Frequency measurements reported by \citealt{mpn+20} are marked with stars. The middle panel shows a piece-wise derivative of the model frequency evolution, averaged over 1-day windows. The  dashed line shows the step changes in the mean frequency derivative included in the model. The lower panel shows the characteristic age derived from this model, which show fluctuations of more than an order of magnitude, highlighting that it is inappropriate to use short-term timing to measure a meaningful characteristic age. In each panel, the maximum-likelihood solution is shown in blue and 256 samples of the Markov chain are overlaid in pink to give an impression of the uncertainty in the model. Vertical bands show the epoch of our observations, with colours indicating where alternate profile shapes ($\mathrm{P_1}$ and $\mathrm{P_2}$) were observed.}
    \label{fig:spin_evolution}
\end{figure*}

The frequency evolution, $\nu(t)$ can be investigated by analytically taking the derivative of the Fourier-basis Gaussian process and adding this to the deterministic spin-down model. Determining $\dot{\nu}(t)$ is more difficult as the power-law red noise has a spectral index less than 4, so the second derivative of the Gaussian process is not a smooth process. Therefore, we approximate the derivative of $\nu(t)$ by computing the average gradient of $\nu(t)$ over windows of one day. This process was repeated on 256 samples from the Markov-chain to give an impression of the allowed range of values from our model. These results are shown in Figure~\ref{fig:spin_evolution}, overlaid with the epochs of our observations. In particular, the figure's lower panel shows the variation of the characteristic age as determined at a given epoch, demonstrating the impracticality of attempting to measure the characteristic age of a magnetar from short term period evolution.

\begin{table}
    \caption{Best-fit values from the MCMC analysis. Upper panel shows results from a model with a step change in $\nu$ at the first epoch. Lower panel is the results without a step change in $\nu$. The last column gives the inferred characteristic age. Values are quoted as the mean of the posterior with a 1-$\sigma$ range in the last digit given in parenthesis.}
    \centering
    \begin{tabular}{c|ccc|cc}
    \hline
    \hline
    Epoch & Start   & $\Delta\nu$ & $\Delta\dot\nu$ &$\dot\nu$ & Age \\
     & MJD & $(\mathrm{\mu Hz})$ & $(\mathrm{\mu Hz^2})$ & $(\mathrm{\mu Hz^2})$ & (yr)  \\
    \hline
    0 & 58922.4      &        &    --   & $-$45 &  260  \\
    1 & 58928.3(1)   & 1.2(5) &  19(5)  & $-$25 & 460 \\
    2 & 58930.9(7)   &        & $-$27(10) & $-$53 &  220\\
    3 & 58935.9(5)   &        &  44(8)  & $-$9  & 1300 \\
    4 & 58947.3(7)   &        & $-$17(2)  & $-$26 &  440\\
    \hline
    0 & 58922.4      &        &    --   & $-$44 &  260  \\
    1 & 58927.5(4)   &        &  21(5)  & $-$23 & 510 \\
    2 & 58930.8(7)   &        &  $-$29(10) & $-$53 &  220\\
    3 & 58935.9(5)   &        &  45(9)  & $-$8  & 1500 \\
    4 & 58947.3(7)   &        & $-$17(2)  & $-$25 &  470\\
   \hline
    \end{tabular}
    \label{timing_table}
\end{table}


\begin{table}
    \caption{Maximum likelihood timing model. Parameters were either fixed (F), solved with linear least-squares (L) or from the MCMC analysis (M). Parameters fitted with least squares are given with an error in the last digit, uncertainties on other values should be taken from Table \ref{timing_table}. The parameters are as reported by \textsc{tempo2} in Barycentric Coordinate Time (TCB) units. $A_\mathrm{red}$ and $\alpha_\mathrm{red}$ are the amplitude at $1 \textrm{yr}^{-1}$ and index of the power-law red noise model as defined in the text.}
    \centering
    \begin{tabular}{l|l|l}
    \hline
    \hline
    Parameter & Value & Type \\
    \hline
        RA (J2000)\dotfill& $18$:$18$:$00.12$ & F\\
        Dec (J2000)\dotfill& $-16$:$07$:$52.8$  & F\\
        $\nu(t_0)$\dotfill& $0.7334047(7)\,\mathrm{Hz}$ & L\\
        $t_0$ (MJD)\dotfill& $58924.0$ & F\\
        $\dot\nu$\dotfill& $-4.45(4)\times10^{-11}\,\mathrm{Hz^2}$ & L\\
        DM\dotfill& $701.8\,\mathrm{cm^{-3}pc}$ & F\\
        Epoch $1$ (MJD)\dotfill& 58928.30 & M \\
        Epoch $2$ (MJD)\dotfill& 58931.00 & M \\
        Epoch $3$ (MJD)\dotfill& 58936.05 & M \\
        Epoch $4$ (MJD)\dotfill& 58947.22 & M \\
        $\Delta\nu_1$\dotfill& $1.45\,\mathrm{\mu Hz}$        & M \\
        $\Delta\dot\nu_1$\dotfill& $17.5\,\mathrm{\mu Hz^2}$  & M \\
        $\Delta\dot\nu_2$\dotfill& $-23.7\,\mathrm{\mu Hz^2}$ & M \\
        $\Delta\dot\nu_3$\dotfill& $42.0\,\mathrm{\mu Hz^2}$ & M \\
        $\Delta\dot\nu_4$\dotfill& $-18.0\,\mathrm{\mu Hz^2}$ & M \\
        \hline
        $\log_{10}(A_\mathrm{red})$ \dotfill & $-5.7(3)$ & M\\
        $\alpha_\mathrm{red}$\dotfill & $3.5(3)$ & M\\
        \hline
    \end{tabular}
    \label{par_table}
\end{table}

\subsection{Radio emission}

\subsubsection{Profile evolution in time and frequency}
\label{sec:profile}

Radio-loud magnetars are known to show drastic change in their radio pulse profile within a few weeks from the onset of the X-ray outburst (e.g. \citealp{crj+07,lld+19}), and this is no different for Swift J1818.0$-$1607. The initial observations with the Effelsberg and Lovell radio telescopes already reported significant changes \citep{kdk+20,rsl+20,cdj+20}. We have seen profile shape changes on timescales from hours to days, while, as shown below, in general only two main types of average pulse shapes are observed which the magnetar is switching in between over the course of our observations. 

\begin{figure}
    \centering
    \includegraphics[scale=0.35]{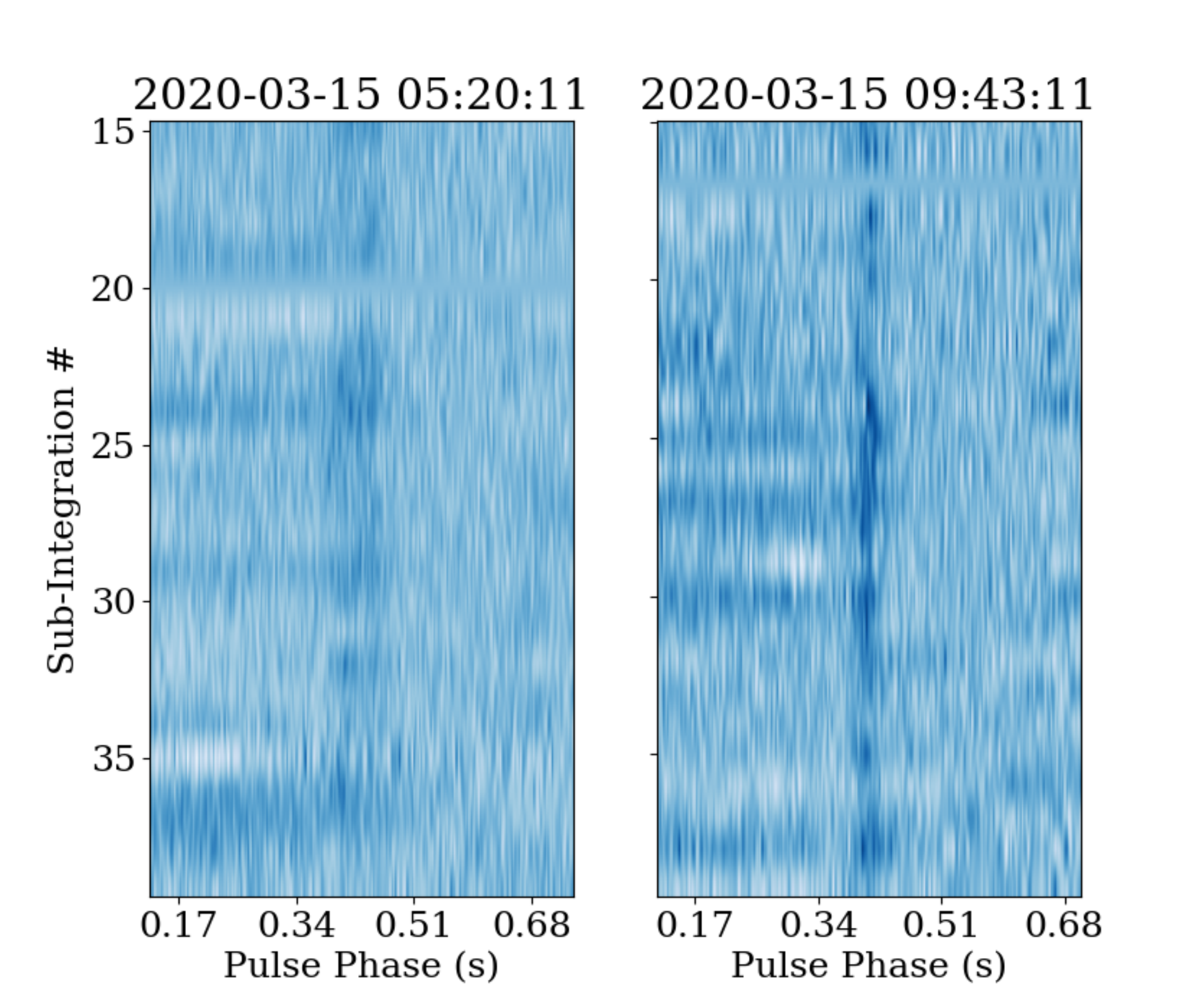}\\
    \includegraphics[scale=0.35]{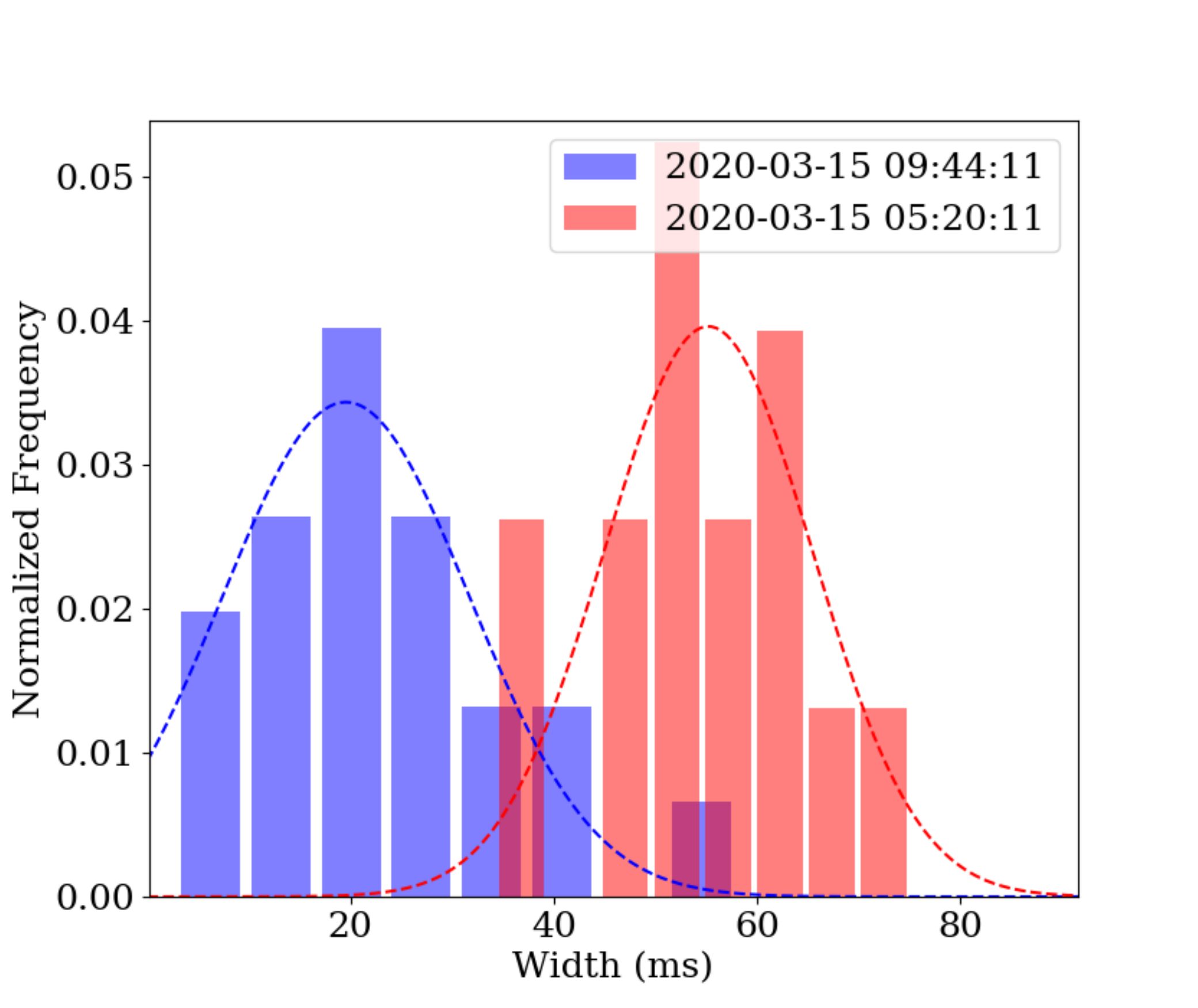}
    \caption{\textbf{Top Panel:} Waterfall plots of 20-second sub-integrations of 1.4 GHz Lovell data clearly showing the width of the pulses before and after the transition. \textbf{Bottom Panel:} Normalized width distribution of 20-second sub-integrations of Swift J1818.0$-$1607 using the data just before the transition (red) and just after the transition to a narrower profile (blue) along with the best fit Gaussian. All times reported here are in UT. }
    \label{fig:width_dist}
\end{figure}

Initial observations with the Lovell Telescope on MJDs 58922 (2020 March 14) and 58923 (2020 March 15) show a wider profile that transitioned into a narrower profile within a span of a few hours (top panel of Figure~\ref{fig:width_dist}) where the average pulses from individual 20\,s sub-integrations switch from wider to narrower profiles between the two observations. To quantify this transition, we fitted a Gaussian function to the pulse profile of each sub-integration. The 20-second length of the sub-integrations was chosen to optimise the signal-to-noise ratio (S/N) for the function fitting and to preserve the best possible time resolution to capture the transition. For each sub-integration, we measured the Full Width Half Maximum (FWHM) of the best-fit Gaussian to the pulse profile. Finally, we generated a histogram of the measured widths of all sub-integrations. We repeated the analysis on two epochs; one where the profile was wider and; the other right after the transition to a narrower pulse profile. The change in profile is evident from the distribution of pulse widths shown in Figure~\ref{fig:width_dist}. 

\begin{figure}
    \centering
    \hspace*{-1cm}
    \includegraphics[scale=0.4]{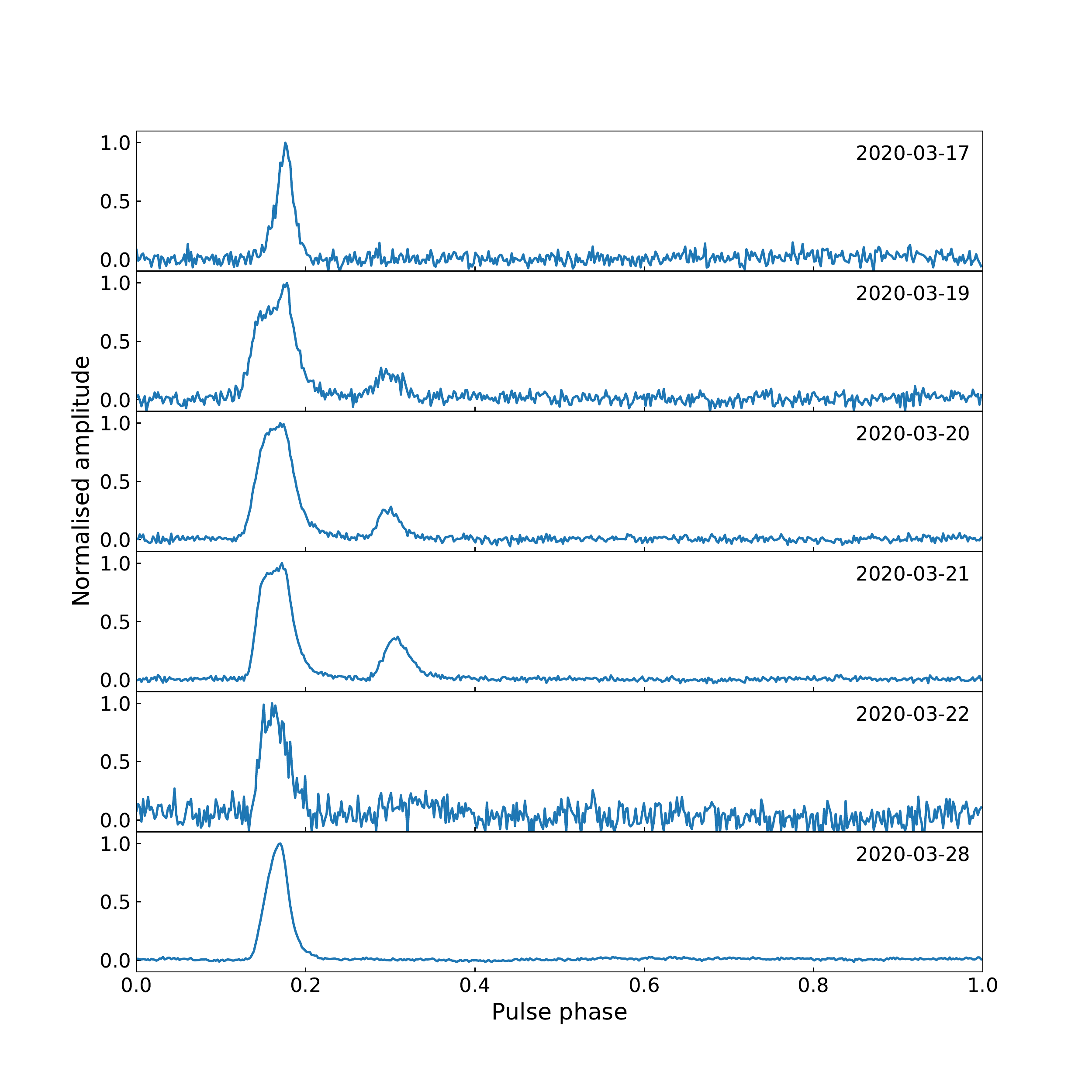}
    \caption{Profile variations over 11 days which span 2020-03-20, the epoch of the first timing event. The amplitudes in each profile were normalized to the peak intensity of the profile.}
    \label{fig:profvar-glitch}
\end{figure}

Further profile changes seem to be associated with the timing events discussed in Section~\ref{sec:timing}. Within a couple of days' window around the epoch MJD 58928 (2020 March 20), i.e.~the first timing event, there is a trend of profile variation which is shown in Figure~\ref{fig:profvar-glitch}. While the pulsar exhibited a single-component structure on MJD 58925
(2020 March 17), a secondary component started to be visible from MJD 58927 (2020 March 19). The leading edge of the main component also started to show up again after fading on MJD 58923 (2020 March 15). The profile stayed approximately the same, until MJD 58930 (2020 March 22), when the second component became barely visible. Then a single-component profile was seen from MJD 58936 (2020 March 28) onward.

The flux density of Swift J1818.0$-$1607 is seen to vary significantly from epoch to epoch, a characteristic typical for other radio-loud magnetars \citep{lev12}. As an extreme case, on  MJD 58935 (2020 March 27), we did not detect the average pulsed emission from the source despite a 3-hr integration time. Instead, we detected only 16 bright individual pulses during those 3 hours. On the contrary, on the next day, MJD 58936 (2020 March 28), a single-component integrated profile was detected with high significance. We discuss single-pulse properties next and refer to the flux density spectrum in Section~\ref{sec:flux}.

\subsubsection{Single-pulse properties}

\begin{figure}
    \centering
    \includegraphics[scale=0.35]{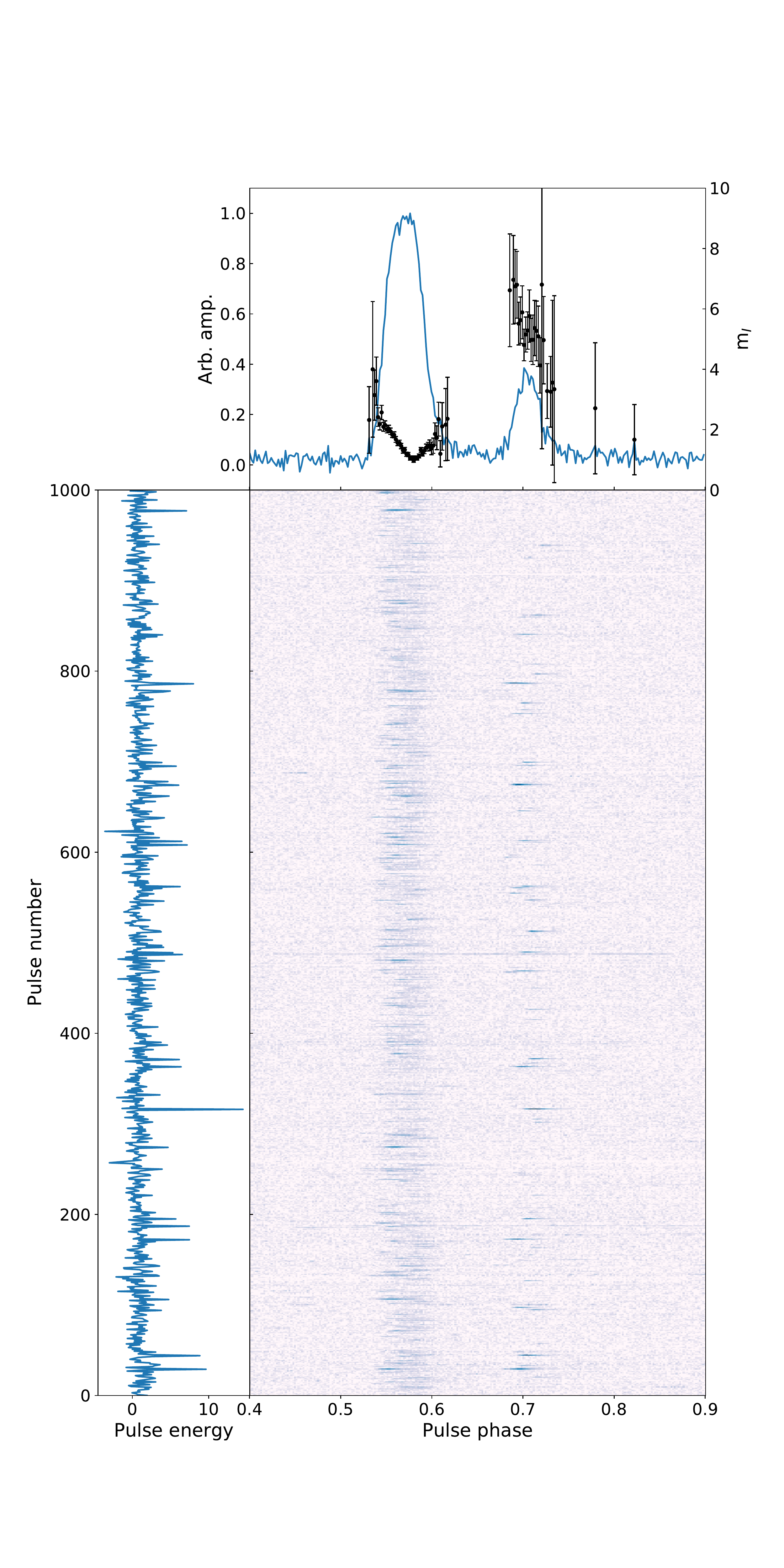}
    \vspace*{-1.5cm}
    \caption{A single-pulse sequence from the 20200320 L-band observation which contains 1,000 rotations. Top right panel: Average profile (solid line) normalised to its peak amplitude, and the longitude-resolved modulation indices (points, without removal of white noise). Bottom left panel: Energy of the single pulses normalized by the mean. Bottom right panel: Colour map of pulse intensity as function of pulse phase and pulse number.}
    \label{fig:waterfall}
\end{figure}

\begin{figure}
    \centering
    \includegraphics[scale=0.35]{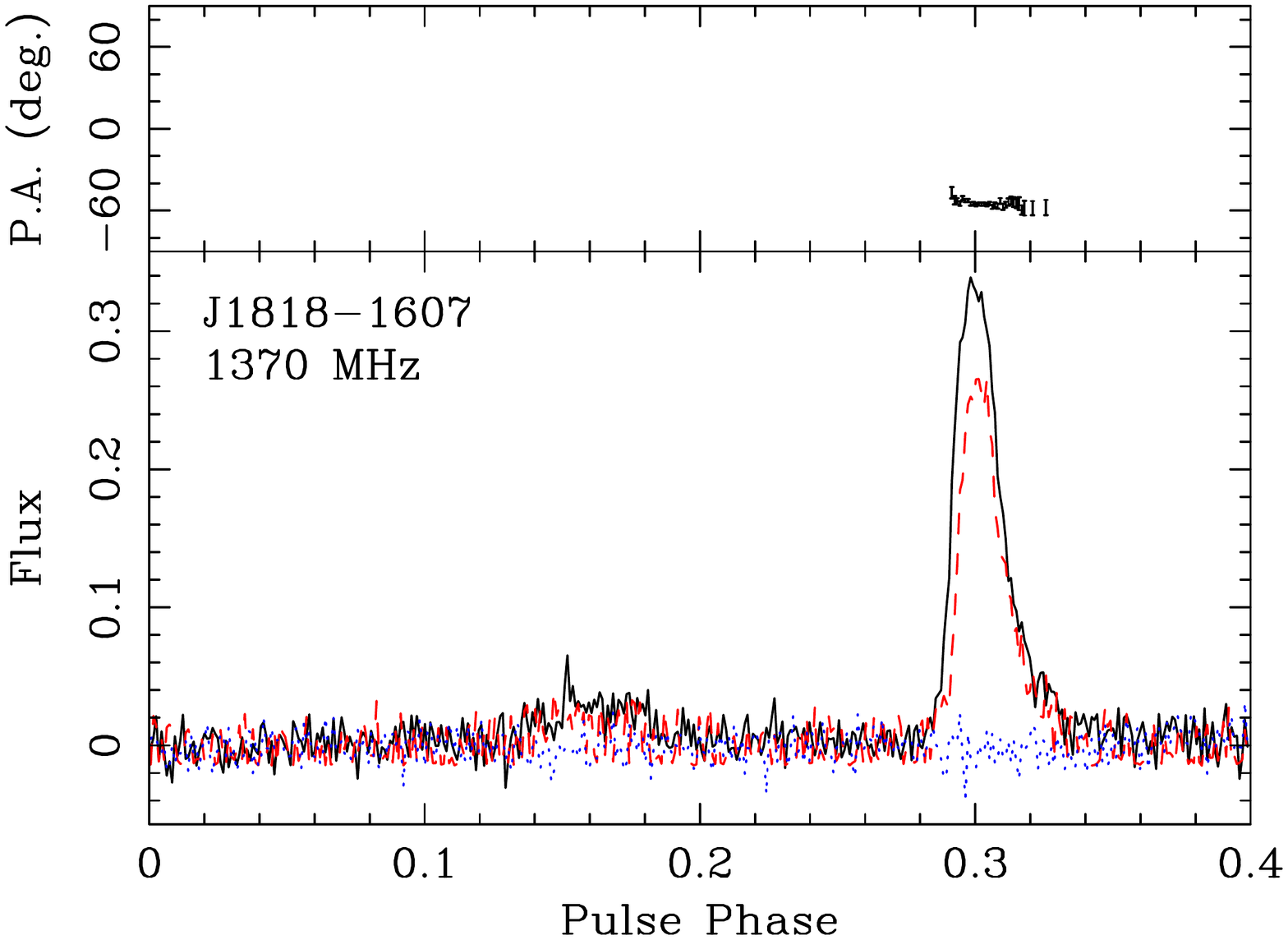}
    \caption{The brightest single pulse from the 20200320 L-band observation. The solid, dashed and dotted line correspond to the total intensity, linear and circular polarisation, respectively. The pulse energy hits approximately 14 times the average, as seen in Figure~\ref{fig:waterfall} (near pulse number 320).}
    \label{fig:sgl}
\end{figure}

The flux density of the magnetar is sufficiently high to enable us to undertake a detailed study of its single pulse properties. It was found that, as in other radio-loud magnetars, Swift J1818.0$-$1607 exhibits sporadic pulse-to-pulse variability. An example of such can be seen in Figure~\ref{fig:waterfall}, where a sequence of 1,000 pulses are shown from the L-band observation taken on MJD 58928 (2020 March 20). As discussed in the section above, there were two components seen in the average pulse profile on that epoch. In the primary component, especially the peak region, the emission is more persistent but that seen in the secondary component is more sporadic. The longitude-resolved modulation index is in general around 2 or below in the region of the primary component, similar to what was seen in XTE J1810–197 at L-band \citep{ssw+09}. The values are in general a factor of 3 larger at the secondary component. This behaviour was also seen at the other epochs when the secondary component was present. The overall single-pulse energy typically fluctuates up to a factor of 5-7 times the average, with however one single pulse seen with energy approximately 14 times the average. The polarisation profile of this pulse is presented in Figure~\ref{fig:sgl}. Most of the pulse energy is concentrated at the phase of the secondary component, with a high degree of linear polarisation. Only a very weak detection can be seen at the phase of the primary component. 

\subsubsection{Polarisation properties} \label{sssec:pol}

\begin{figure*}
\includegraphics[scale=0.38]{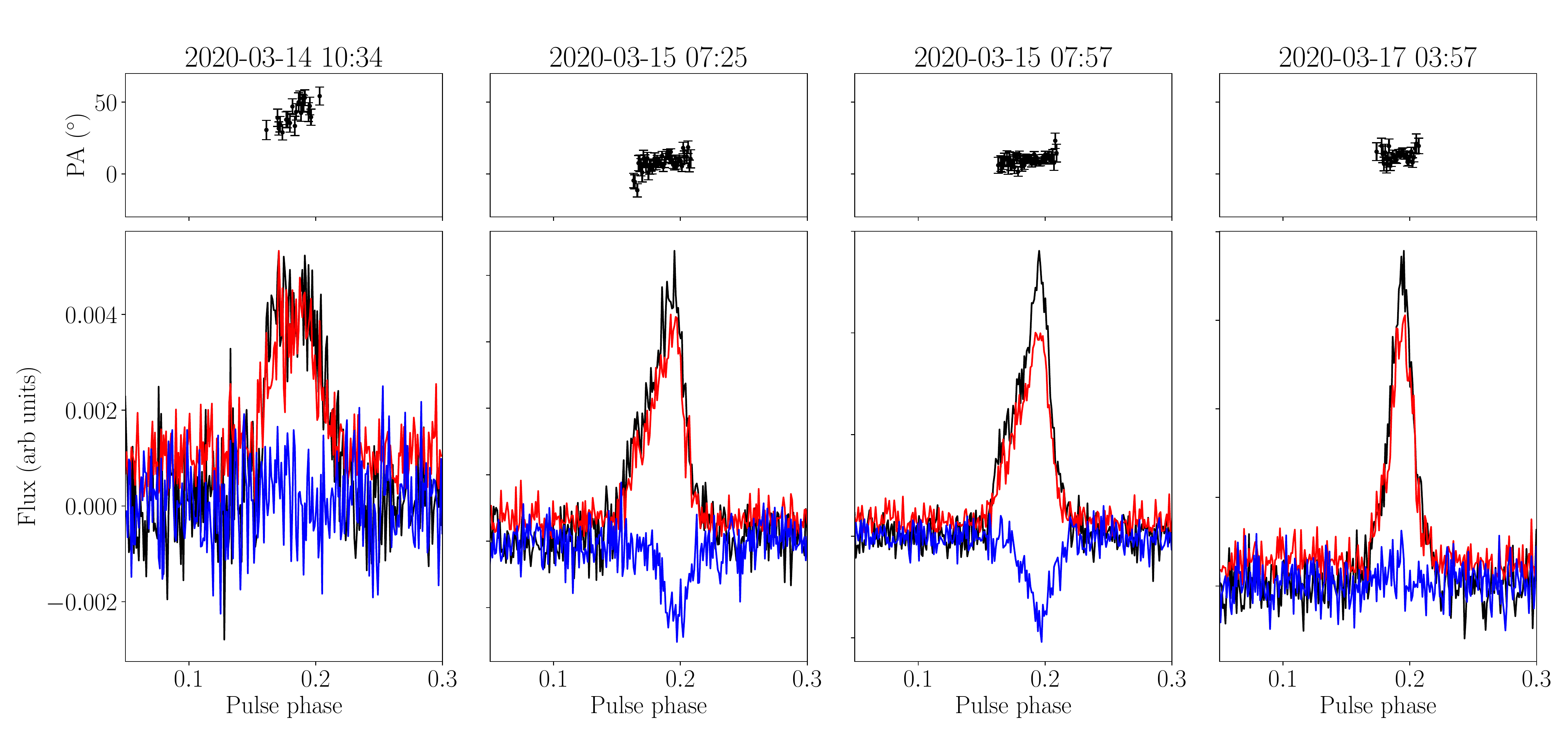}\\
\includegraphics[scale=0.38]{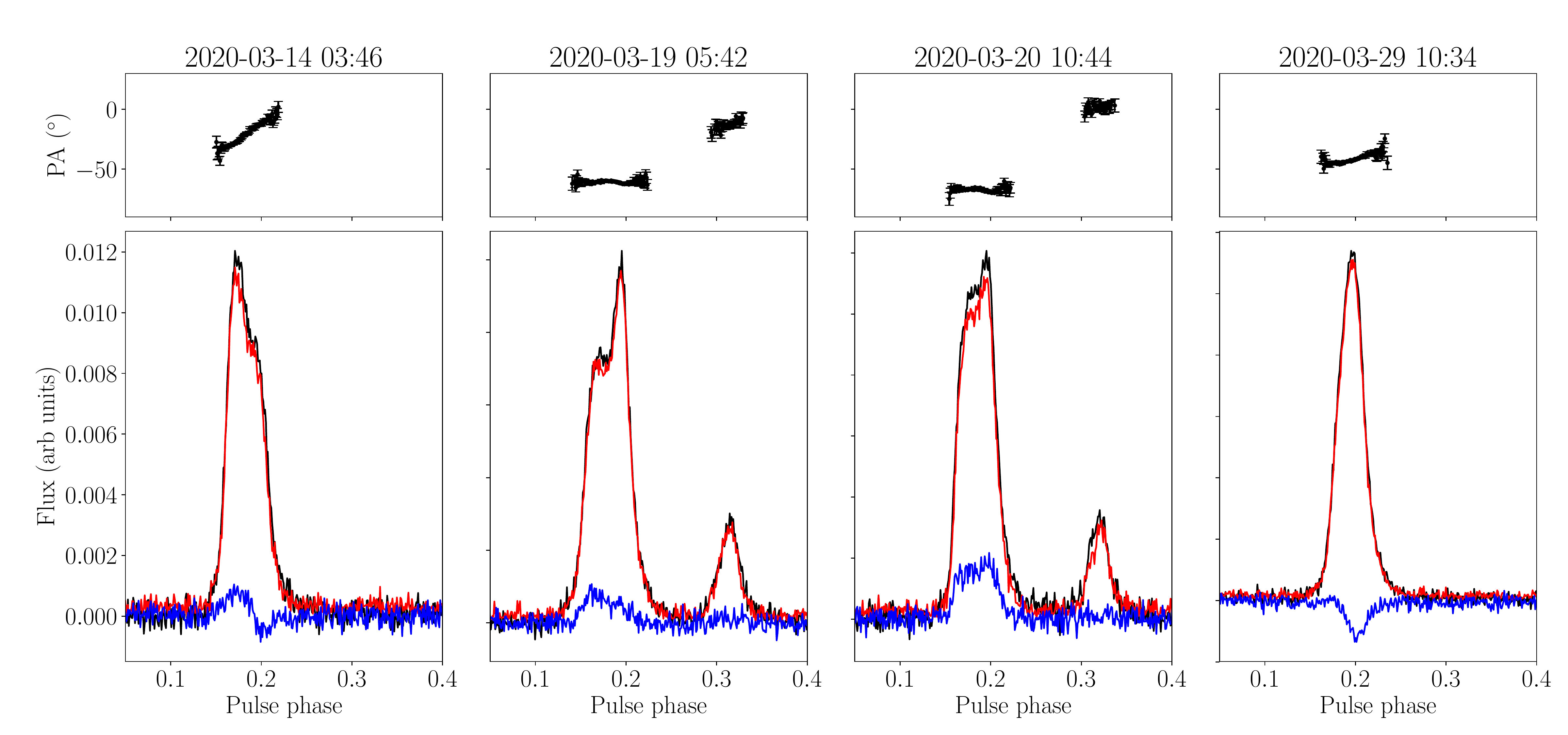}
\caption{Average pulse profile of Swift J1818.0$-$1607 showing total intensity (solid black), linear polarization (solid red) and circular polarization (solid blue) from JBO (Top Row) and Effelsberg (Bottom Row). A flattening of the PA during the profile transition is clearly evident in the JBO data. The data have been corrected for Faraday rotation using an RM value of $+1442(3)$ rad m$^{-2}$.} 
\label{fig:1818_poln}
\end{figure*}


The first polarisation profile of Swift J1818.0$-$1607 was presented by \cite{kdk+20}. Follow-up observations at L-band, as shown in Figure~\ref{fig:1818_poln}, have consistently measured a very high degree of polarisation in the pulse profile which is dominated by nearly 100\% linearly polarised emission. The associated position angle (PA) is mostly flat, but occasionally shows a small rising slope. When the second, trailing profile component is visible, it also shows a mostly flat PA that is offset from the PA in the first component by about 60\,deg. In contrast to the high degree of linear polarisation, the profiles typically show a small degree of circular polarisation which sometimes changes handedness near the pulse centre.

\subsubsection{Flux density timeline and spectrum}
\label{sec:flux}

\begin{figure}
\centering
    \includegraphics[width=\columnwidth]{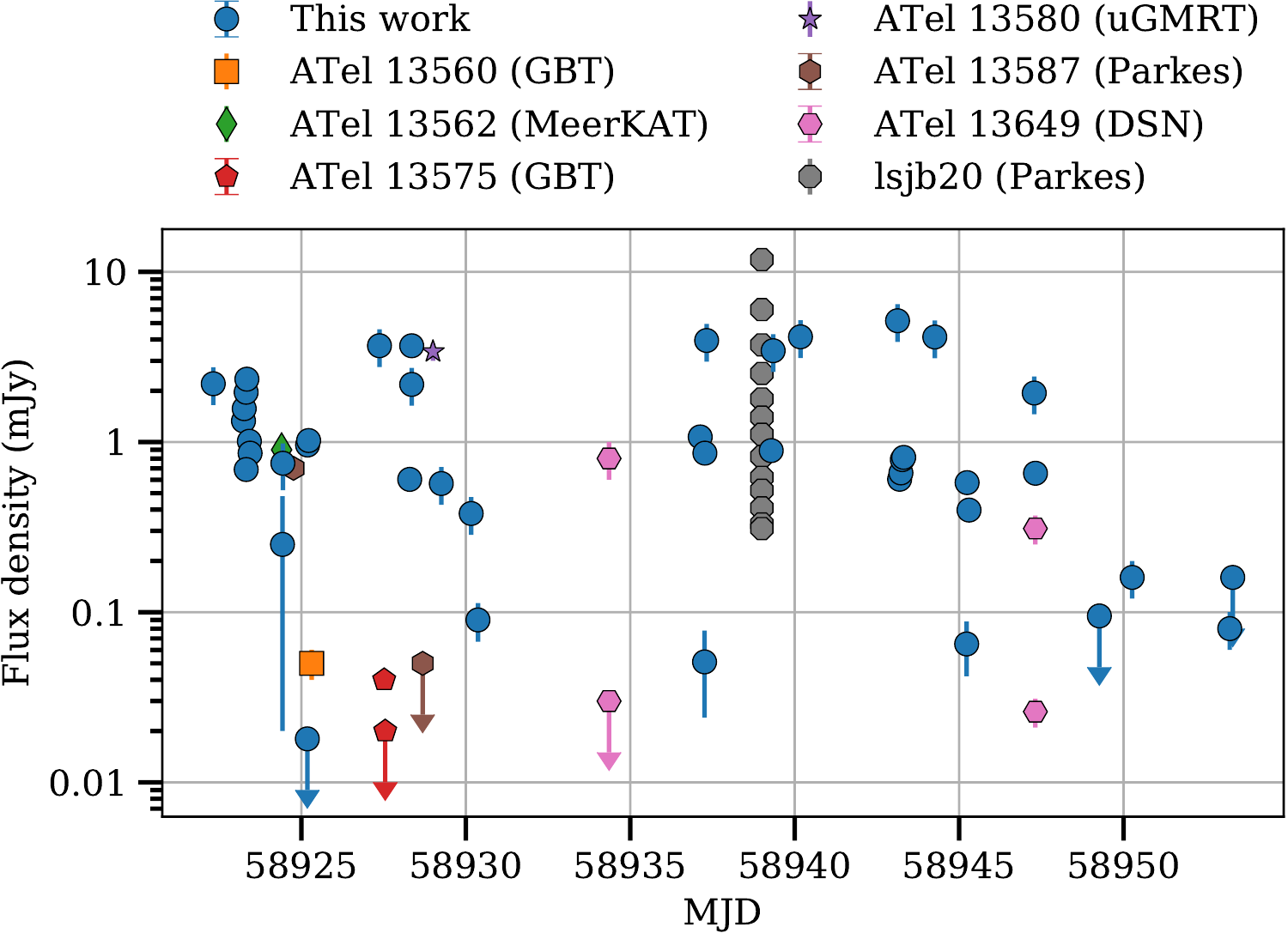}
    \includegraphics[width=\columnwidth]{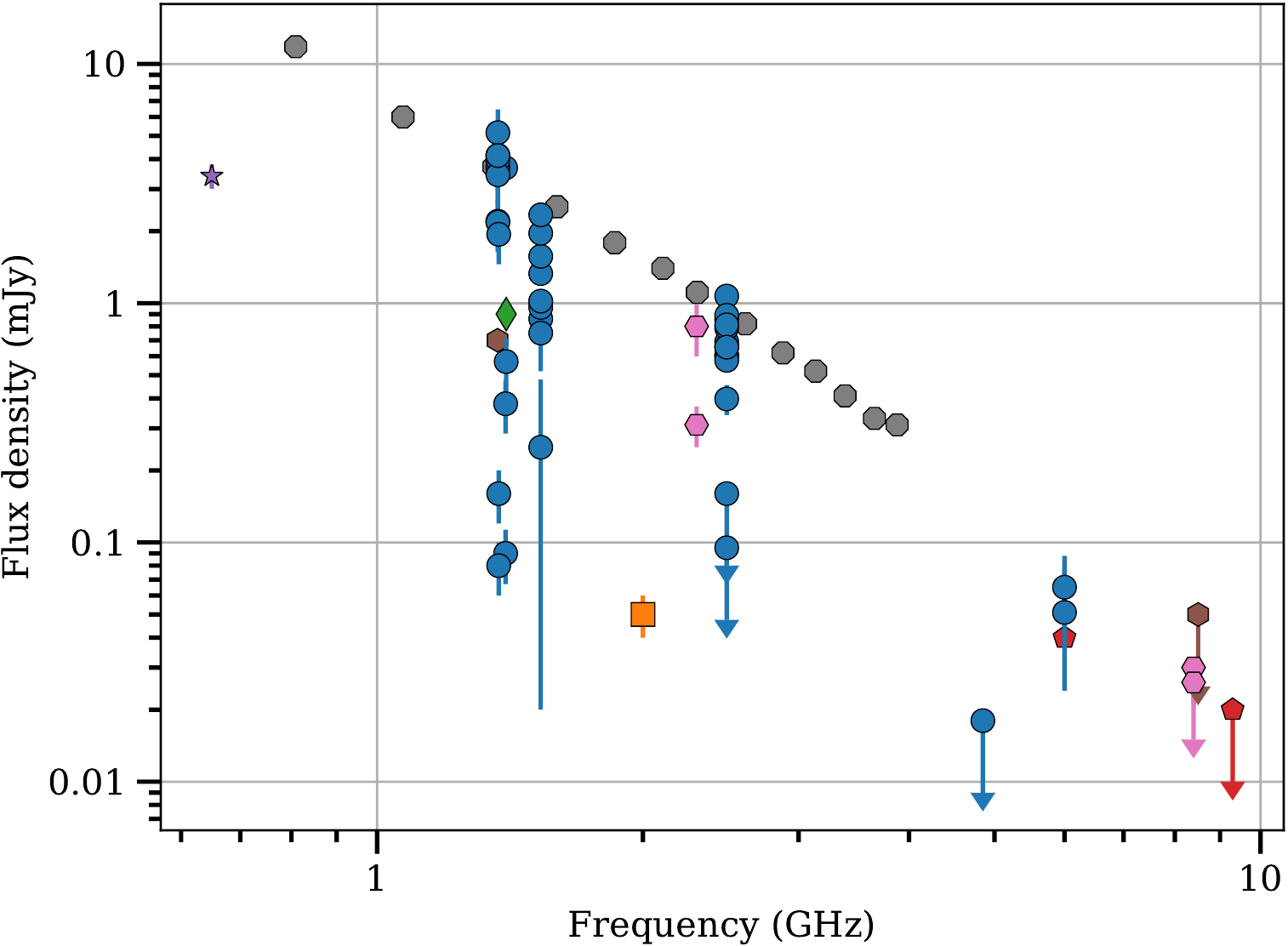}
    \includegraphics[width=\columnwidth]{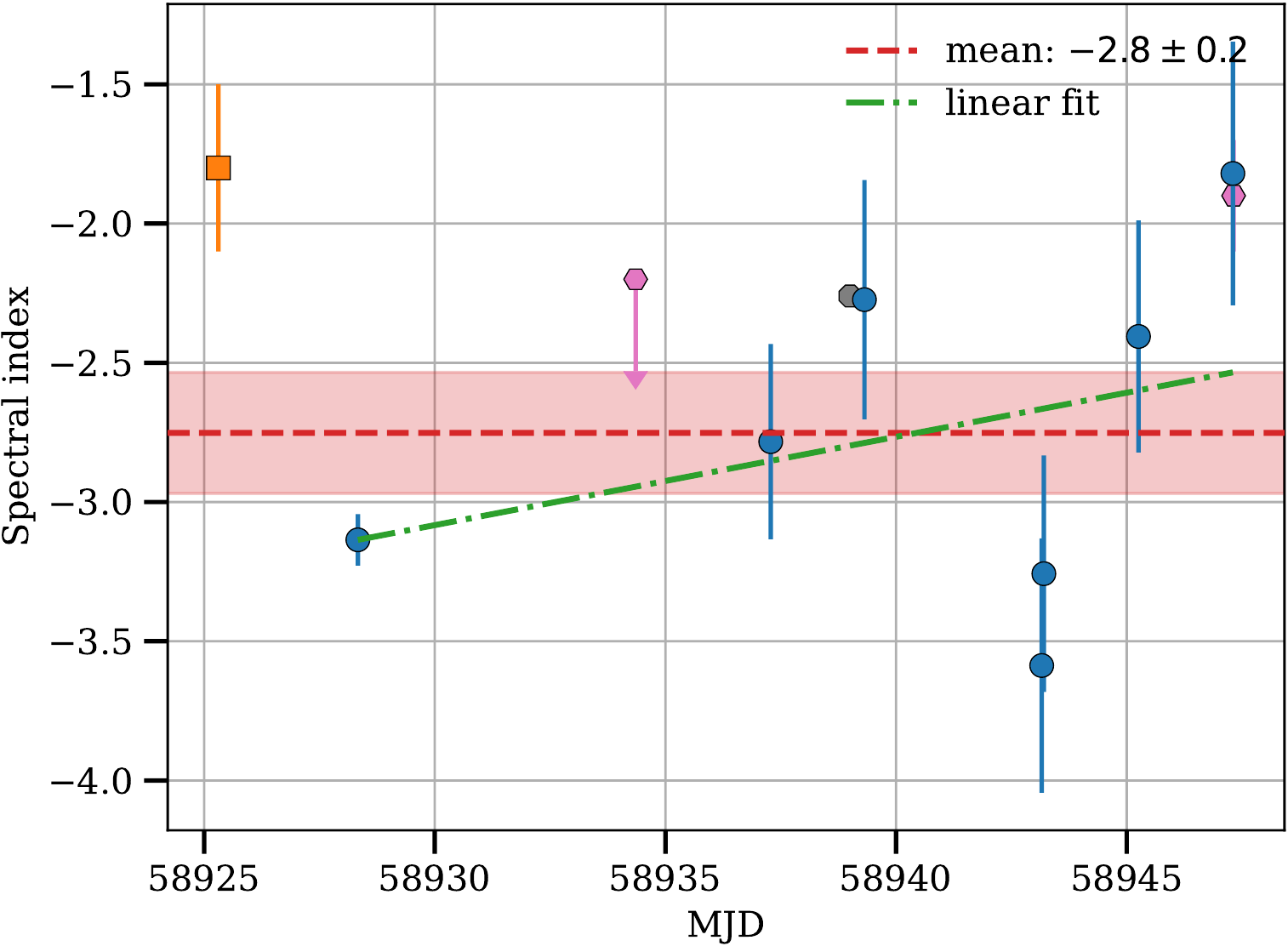}
    \caption{Top panel: Timeline of the pulse-averaged flux density of Swift~J1818.0$-$1607 at multiple radio frequencies, based on observations with the Effelsberg and Lovell telescopes, together with a selection of measurements reported in the literature. The markers reference the same data points in all panels. Middle panel: The same multiple-epoch data as above, but plotted against frequency. Bottom panel: Spectral indices along with their $1-\sigma$ uncertainties from both the literature and our work, the mean spectral index computed over all epochs, and the best-fitted linear function to the spectral indices in time. The spectral indices were obtained from fitting a simple power-law model to quasi-simultaneous band-integrated data from the Effelsberg telescope at two or three frequency bands ($\sim 1.5$ to $\sim 6.8$~GHz frequency coverage).}
\label{fig:spectrum}
\end{figure}

We estimated the pulse-averaged flux densities of the Lovell and Effelsberg Telescope multi-frequency data to characterise the magnetar's temporal evolution in flux density and its radio spectrum from quasi-simultaneous observations. For the data from the Lovell Telescope, we used a previously determined analytical parameterisation for its system-equivalent flux density (SEFD) as a function of elevation to reference our measurements to a known flux density scale. Observations of high-DM pulsars with well-known absolute calibrated flux densities at L-band (from \citealt{jvkb+18}), that were performed close in time to the magnetar observations, were used to ascertain the fidelity of our calibration method. The Effelsberg data are from observations at L, S and C-band, for which the calibration methods differ. Both the S and C-band data were referenced to an absolute flux density scale based on observations of a primary flux density calibrator, the planetary nebula NGC~7027. We employed a standard parameterisation for its radio spectrum. For those data, as well as the ones from the Lovell Telescope, we visually identified on-pulse phase gates per epoch that included all pulse profile components and measured the band-integrated pulse-averaged flux densities after baseline subtraction. The uncertainties were derived from the RMS of the off-pulse regions. The L-band measurements were based on the radiometer equation \citep{dew85}, extrapolated sky temperatures to the reference frequencies \citep{1982Haslam, law87} and known receiver performance parameters\footnote{\url{https://eff100mwiki.mpifr-bonn.mpg.de}}. We assumed a 25~per cent uncertainty for those measurements.

In Fig.~\ref{fig:spectrum}, we show a timeline of our flux density measurements together with a selection of data points and upper limits at various radio frequencies reported in the literature. Where literature data are available at epochs close to our measurements (and frequencies), they generally agree well. The middle panel shows the same multi-epoch data set in a spectrum. Finally, the bottom panel presents the best-fitting spectral indices from fitting a simple power law of the form $S (\nu) \propto \nu^{\alpha}$, where $S$ is the pulse-averaged flux density at frequency $\nu$ and $\alpha$ is the spectral index, to quasi-contemporaneous band-integrated data from the Effelsberg telescope obtained at two or three frequency bands. While it would be possible to split some of the data into frequency sub-bands, this is referred to future work. Our spectral fits cover the frequency ranges L to S ($\sim 1.5$~GHz), S to C ($\sim 5.7$~GHz), or the full L to C-band range ($\sim 6.8$~GHz). In total, we derive spectral indices at seven epochs, and the vast majority of those observations are nearly contiguous in time. We performed robust and uncertainty-weighted parameter estimation using MCMC techniques, for which we employed the \textsc{emcee} software. The start parameters for the MCMC runs were set based on initial maximum likelihood fits. We measured spectral indices that vary between $-3.6$ and $-1.8$, with a mean spectral index over all epochs of $-2.8 \pm 0.2$, where the uncertainty is the standard error of the mean. Our measurements agree well with estimates from the literature obtained close to our observing epochs, e.g.\ those from the Parkes telescope on MJD~58939 (2020 March 31) between about 0.7 to 4~GHz \citep{lsjb2020} and the one from the Deep Space Network on MJD~58947 (2020 April 08) between 2.3 and 8.4~GHz \citep{mpn+20}, which further reassured us of the fidelity of our calibration methods. The magnetar's radio spectrum was therefore surprisingly steep at the times of measurement and showed a significant amount of variability, as can be seen in the early data before MJD~58925 in particular. Additionally, from purely visual inspection it seemed that the spectrum became flatter over time (bottom panel of Fig.~\ref{fig:spectrum}). To test that hypothesis, we fit both a constant and a linear function to the spectral index time series using the same techniques as before and performed model selection using the Akaike information criterion (e.g.~\citealt{2010Burnham}). We found that there is strong statistical evidence for a spectral flattening over time. The slope of the best-fitting linear function has a formal statistical significance in excess of $6\,\sigma$, and the probability that the linear function is the best-fitting one among the two models tested is about 70~per cent. We conclude that we saw an initial transition of the source to a more magnetar-like, i.e.\ flatter, radio spectrum over a few weeks. Interestingly, \citet{majidinversion} recently reported the measurement of an inverted spectrum of the magnetar with spectral index of +0.3 between 2.3 and 8.4~GHz in mid July 2020.

\section{Discussions \& Conclusions}

We have studied the newly discovered magnetar Swift J1818.0$-$1607 with very high cadence for more than one month after its activation. The magnetar-nature of the source is confirmed by a number of observed traits similar to the properties of other radio-loud magnetars. 

The spin-down measurements imply a large magnetic field strength, a young age and a large spin-down luminosity. However, we also show that the spin-down derivative changes by a factor of a few within days and weeks. Hence, the inferred characteristic age of the magnetar varies accordingly over the course of one month. As seen from Figure~\ref{fig:spin_evolution}, the lowest inferred characteristic age is about 250 years, consistent with the age inferred first by \citet{cdj+20} and later \citet{esposito20}. However, only two weeks later, one determines a characteristic age of about 2000 years which is significantly larger. 

\begin{figure}
    \centering
    \includegraphics[scale=0.7]{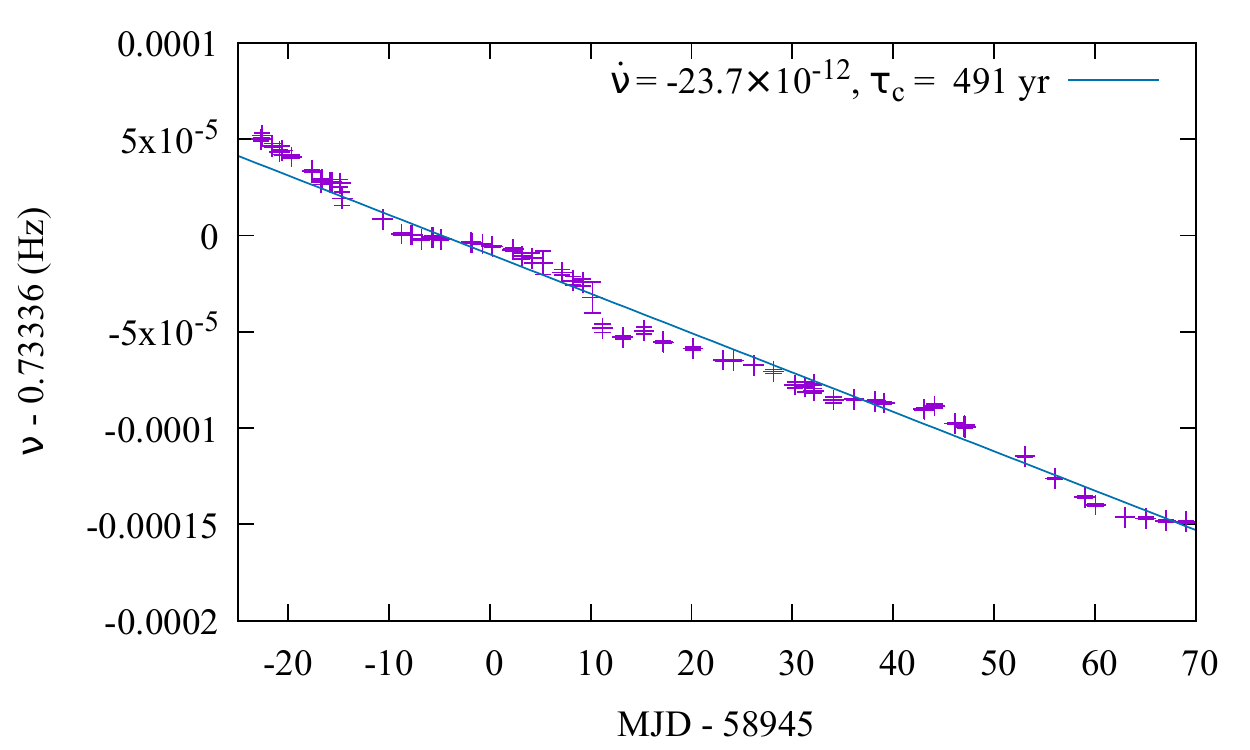}
    \caption{Measurement of rotational frequency from individual epochs, together with a linear fit to the overall variation trend. The characteristic age was calculated with the $\dot{\nu}$ obtained from the fit. }
    \label{fig:F0s}
\end{figure}

\begin{figure}
    \centering
    \includegraphics[scale=0.4]{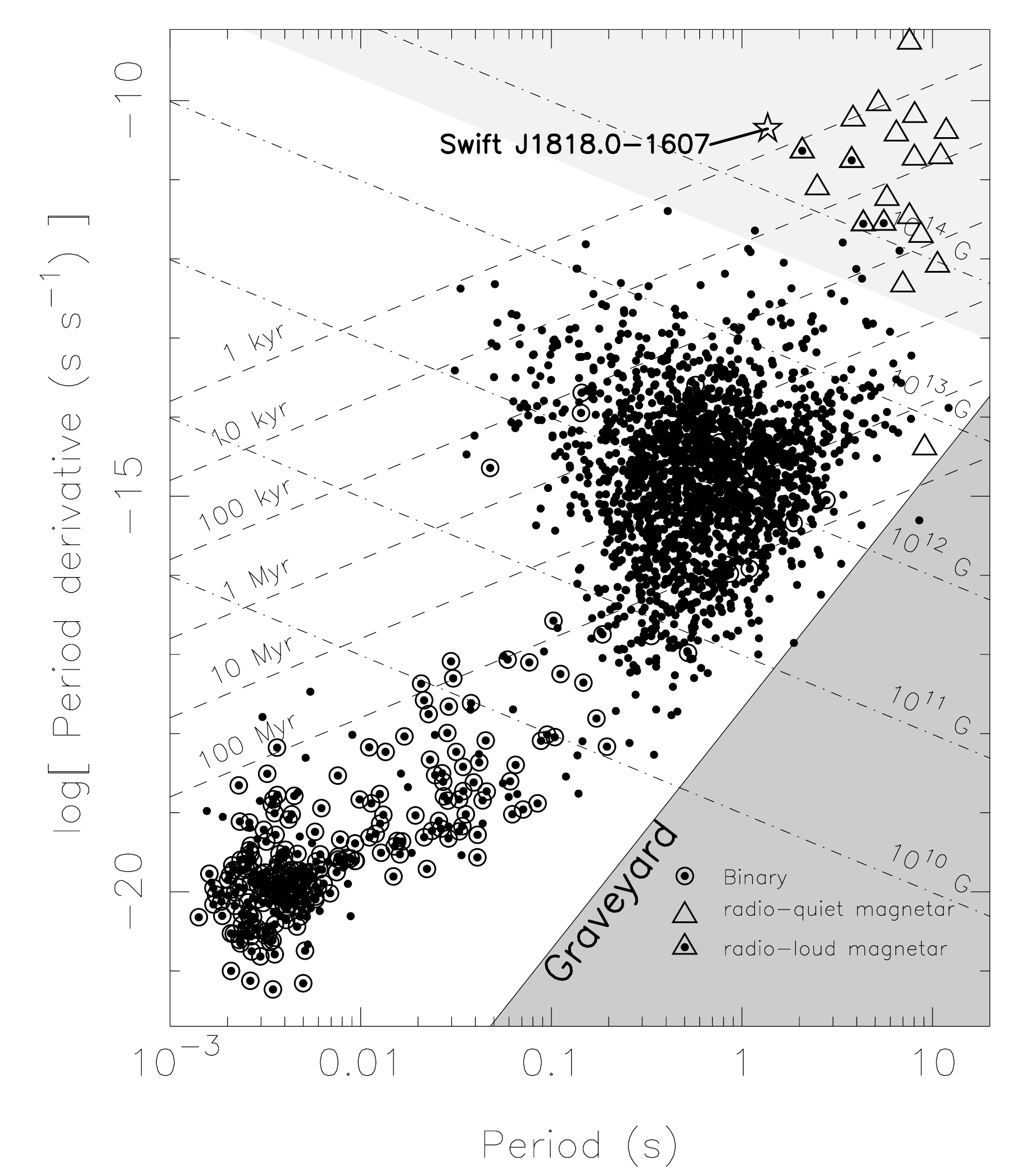}
    \caption{$P$-$\dot{P}$ diagram with the position of Swift J1818.0$-$1607 marked. The light-gray area in the top-right marks inferred magnetic field values above the quantum critical field.}
    \label{fig:ppdot}
\end{figure}

To obtain a more reliable estimate of the overall spin-down rate, we measured the rotational frequency from each individual observations over a longer period of time and in Figure~\ref{fig:F0s}, we present data obtained for a span of nearly 100 days since activation. Fitting a linear slope to these data which averages over the short-term variations in spin frequency, we measure a spin frequency derivative of $\dot\nu = -24\times 10^{-12}$\,$\rm Hz^{2}$, implying a characteristic age of about 500 years. We consider this a more reliable characteristic age estimate. Nevertheless, one should be cautious on whether the calculated characteristic age (based on $\tau = P/2\dot{P}$) reflects the true age of the magnetar, because the braking index can well be deviated from 3 or even time-varying \cite[e.g., ][]{jk17}, and the birth period could be significantly less than the current period. In particular, a braking index less than 3, which has already been observed in a small number of magnetars \citep{glw+16}, could well lead to an underestimation of the true age while using the value of the characteristic age. 
Such caveats also apply to previous claims of the age of the magnetar stated in \cite{cdj+20} and \cite{esposito20}.
In fact, there is no reported coincident supernova remnant that could help to further constrain the age of the magnetar \citep{green19, lsjb2020}, though a non-detection may suggest a true age of a higher value.
The combination of the spin period and spin-down rate locates Swift J1818.0$-$1607 at a position in the $P$-$\dot{P}$-diagram (see Fig.~\ref{fig:ppdot}) which expands the current boundary of magnetars. In fact, it would be the fastest spinning magnetar known if PSR~J1846$-$0258 is considered to be from a distinct pulsar population as discussed in \citep{ggg+08,lnk+11,khd18}.

Similar to other magnetars \citep[cf.][]{kb17}, Swift J1818.0$-$1607 also shows a number of distinct timing events. Due to the high-cadence of our observations, we identify four distinct events in the first month after the outburst. All timing events are characterised by a change in spin-down rate, whereas the first event also shows a change in spin frequency. The first event coincides with a period of time when, first, the profile changes significantly over a short timescale of hours
(see Figure~\ref{fig:width_dist}) and then develops a distinct second profile component (see Figure~\ref{fig:profvar-glitch}), which is also characterised by a number of very strong single pulses (see Figures~\ref{fig:waterfall} \& \ref{fig:sgl}). Such a relationship between changes in the observed pulse profile, and presumably magnetospheric structure, and spin-evolution is also well known from `normal' radio pulsars \citep{lyne2010}. It is worth noting that although our model selection strongly prefers a model containing 4 step changes in $\dot{\nu}$ in addition to power-law red noise, we caution that this does not rule out the general class of smooth models of $\dot{\nu}$ variations, and that as seen in Figure 2(b), there is significant smooth variation in $\dot{\nu}$ at a scale only slightly below that of the discrete events.

The rapid onset of radio emission, and the high cadence timing observations we have performed, have allowed us to study the coherent timing behaviour of this source shortly after its outburst. The bulk timing behaviour shows variations which are similar (in a relative sense) to those seen in the X-rays in the early days after the outburst from XTE J1810$-$197 \citep{iam+04} but with more detail. These variations may also provide vital information on what is happening to the underlying neutron star immediately after it returns to quiescence.

We note that the exact epoch of the first timing event is estimated to coincide with Effelsberg observations at 2.7 GHz. We have investigated the timing behaviour as well as the emission properties around the estimated epoch, but the relatively low strength of the magnetar emission at this frequency on this day prevents us from studying information based on single pulses, which may have revealed magnetospheric changes \citep[cf.][]{velaglitch}. Hence, given the available time resolution and signal strength, we do not detect significant changes.

Similarly to other radio-loud magnetars, Swift J1818.0$-$1607, shows very highly polarised radio emission and a pulse profile that changes shape, although significantly less dramatically than other magnetars. The modulation index determined by the pulse-to-pulse fluctuation is high, and the second profile component often shows the `spiky' narrow single pulses also seen in other sources. The PA swing shows variations as in other magnetars, but is mostly flat, with an offset between the first and second component when the latter is observed.
If the PA swing is interpreted geometrically, one would conclude that our line-of-sight grazes the emission beam at its edges, away from the magnetic axis. If one additionally assumes a frequency-dependent emission height, a shrinkage of the beam could explain the unusually steep flux density spectrum, as the emission would recede from our view at higher radio frequencies. The steep spectrum is indeed in stark contrast with measurements from other radio-loud transient magnetars that show high radio emission variability and generally flat ($\alpha \geq -0.5$; e.g.~\citealt{2007CamiloC, keith2011, tor15, dai2019}) or even slightly inverted spectra up to millimetre wavelengths \citep{tor17}. It is interesting to note that FRBs also tend to show flat position angle swings \citep[e.g.][]{FRBpol}, and indeed the single pulse shown in Figure~\ref{fig:sgl} or the average profiles shown in Figure~\ref{fig:1818_poln} do resemble observed FRB pulse shapes. Given the short rotational period by magnetar-standards (Fig.~\ref{fig:ppdot}), this source may provide yet another indication that both source populations are related.

%
%
To summarise, we have demonstrated the importance of complementing X-ray with rapid radio follow-up observations for studying the initial state of magnetars after their activation.
In order to provide an unbiased view on the spin-evolution of these young neutron stars, a high cadence is required. A study of the associated radio emission properties does not only help to unambiguously identify sources as magnetars with their distinct features, but may also connect the spin behaviour to magnetospheric processes.

\section*{Acknowledgements}

FJ, KR and BS acknowledge funding from the European Research Council (ERC) under the European Union's Horizon 2020 research and innovation programme (grant agreement No.~694745). GD, RK, KL, PT acknowledge the financial support by the European Research Council for the ERC Synergy Grant BlackHoleCam under contract no. 610058. Pulsar research at Jodrell Bank Centre for Astrophysics and Jodrell Bank Observatory is supported by a consolidated grant from the UK Science and Technology Facilities Council (STFC). This publication is based on observations with the 100-m telescope of the MPIfR (Max-Planck-Institut für Radioastronomie) at Effelsberg. We thank Alex Kraus for scheduling our observations at the 100-m telescope so flexibly.
The Nan\c cay Radio Observatory is operated by the Paris Observatory,
associated with the French Centre National de la Recherche Scientifique (CNRS).

\section*{Data Availability}
The data underlying this article will be shared on reasonable request to the corresponding author.




\bibliographystyle{mnras}
\bibliography{1818} 





\bsp	
\label{lastpage}
\end{document}